\documentstyle[prl,aps,epsfig] {revtex}
\begin{document}
\draft
\bibliographystyle{alpha}
%
%
%
\title{
First measurement of the Gerasimov-Drell-Hearn integral
for \mbox{\boldmath$^1$}H from 200 to 800 MeV
}
%
%
\author{J.~Ahrens$^9$,
S.~Altieri$^{15,16}$,
J.R.M.~Annand$^6$,
G.~Anton$^3$,
H.-J.~Arends$^9$\footnote{corresponding author: e-mail address
arends@kph.uni-mainz.de},
K.~Aulenbacher$^9$,
R.~Beck$^9$,
C.~Bradtke$^1$,
A.~Braghieri$^{15}$,
N.~Degrande$^{4}$,
N.~d'Hose$^{5}$,
H.~Dutz$^2$,
S.~Goertz$^1$,
P.~Grabmayr$^{17}$,
K.~Hansen$^8$,
J.~Harmsen$^1$,
D.~von~Harrach$^9$,
S.~Hasegawa$^{13}$,
T.~Hasegawa$^{11}$,
E.~Heid$^9$,
K.~Helbing$^3$,
H.~Holvoet$^4$,
L.~Van~Hoorebeke$^4$,
N.~Horikawa$^{14}$,
T.~Iwata$^{13}$,
P.~Jennewein$^9$,
T.~Kageya$^{14}$,
B.~Kiel$^3$,
F.~Klein$^2$,
R.~Kondratiev$^{12}$,
K.~Kossert$^7$,
J.~Krimmer$^{17}$,
M.~Lang$^9$,
B.~Lannoy$^4$,
R.~Leukel$^9$,
V.~Lisin$^{12}$,
T.~Matsuda$^{11}$,
J.C.~McGeorge$^6$,
A.~Meier$^1$,
D.~Menze$^2$,
W.~Meyer$^1$,
T.~Michel$^3$,
J.~Naumann$^3$,
R.O.~Owens$^6$,
A.~Panzeri$^{15,16}$,
P.~Pedroni$^{15}$,
T.~Pinelli$^{15,16}$,
I.~Preobrajenski$^{9,12}$,
E.~Radtke$^1$,
E.~Reichert$^{10}$,
G.~Reicherz$^1$,
Ch.~Rohlof$^2$,
D.~Ryckbosch$^4$,
F.~Sadiq$^6$,
M.~Sauer$^{17}$,
B.~Schoch$^2$,
M.~Schumacher$^7$,
B.~Seitz$^7$\footnote{present address: DESY/HERMES (Hamburg)},
T.~Speckner$^3$,
M.~Steigerwald$^9$,
N.~Takabayashi$^{13}$,
G.~Tamas$^9$,
A.~Thomas$^9$,
R.~van de Vyver$^4$,
A.~Wakai$^{14}$,
W.~Weihofen$^7$,
F.~Wissmann$^7$,
F.~Zapadtka$^7$,
G.~Zeitler$^3$\\
(GDH- and A2- Collaborations)}
%
\address{$^1$  Inst. f\"ur Experimentalphysik, Ruhr-Universit\"at Bochum,
  D-44801 Bochum, Germany}
\address{$^2$  Physikalisches Institut, Universit\"at Bonn, D-53115 Bonn, Germany}
\address{$^3$  Physikalisches Institut, Universit\"at Erlangen-N\"urnberg, D-91058 Erlangen, Germany}
\address{$^4$ Subatomaire en Stralingsfysica, Universiteit Gent, B-9000 Gent, Belgium}
\address{$^5$  CEA Saclay, DSM/DAPNIA/SPhN, F-91191 Gif-sur-Yvette Cedex, France}
\address{$^6$  Department of Physics \& Astronomy, University of Glasgow, U.K.}
\address{$^7$  II.Physikalisches Institut, Universit\"at G\"ottingen, D-37073
G\"ottingen, Germany}
\address{$^8$  Department of Physics, University of Lund, Lund, Sweden}
\address{$^9$  Institut f\"ur Kernphysik, Universit\"at Mainz, D-55099 Mainz, Germany}
\address{$^{10}$  Institut f\"ur Physik, Universit\"at Mainz, D-55099 Mainz, Germany}
\address{$^{11}$  Faculty of Engineering, Miyazaki University, Miyazaki, Japan}
\address{$^{12}$  INR, Academy of Science, Moscow, Russia}
\address{$^{13}$  Department of Physics, Nagoya University,  Chikusa-ku, Nagoya, Japan}
\address{$^{14}$  CIRSE, Nagoya University,  Chikusa-ku, Nagoya, Japan}
\address{$^{15}$  INFN, Sezione di Pavia, I-27100 Pavia, Italy}
\address{$^{16}$ Dipartimento di Fisica Nucleare e Teorica, Universit\`a di Pavia, I-27100 Pavia, Italy}
\address{$^{17}$  Physikalisches Institut, Universit\"at T\"ubingen, D-72076
  T\"ubingen, Germany}
%
%
\date{\today}
\maketitle
\begin{abstract}
A direct measurement of the helicity dependence of the total
photoabsorption cross section on the proton was carried out at
MAMI (Mainz) in the energy range 200 $<$ E$_\gamma < 800$~MeV.
The experiment used a 4$\pi$ detection system, a circularly
polarized tagged photon beam and a frozen spin target.
The contributions to  the Gerasimov-Drell-Hearn sum rule and to the
forward spin polarizability $\gamma_0$ determined from the data  are
 226$\pm 5 (stat)\pm 12(sys)$ $\mu$b and
-187$\pm 8 (stat)\pm 10(sys)$ $10^{-6}$fm$^4$,  respectively, for 200 $<$ E$_\gamma < 800$~MeV.
\end{abstract}


%
\section{Introduction}
In a recent paper~\cite{first} we have published the first data on
the helicity dependence of the $\gamma p \rightarrow N \pi$
reactions in the energy range from 200 to 450 MeV.
In the present paper we present the
helicity dependence of the total photoabsorption cross
section on the proton in the photon
 energy range from 200 to 800 MeV.
These data give, in particular,
important experimental information about the Gerasimov-Drell-Hearn
(GDH) sum rule and the forward spin polarizability $\gamma_0$.

The GDH sum rule relates the total absorption cross section of
circularly polarized photons on longitudinally polarized nucleons
to the static properties of the nucleon~\cite{GDH}. The two
relative spin configurations, parallel or antiparallel, determine
the two absorption cross sections $\sigma_{3/2}$ and
$\sigma_{1/2}$. The integral over the photon energy $\nu$ of the
difference of these two cross sections, weighted by the inverse
of $\nu$, is related to the mass, $M$, and anomalous
magnetic moment, $\kappa$, of the nucleon as follows:
\begin{equation}\label{eq1}
\int_{\nu_0}^\infty (\sigma_{3/2} - \sigma_{1/2})\,{\hbox{d}\nu \over \nu}
= \frac{2\pi^2 \alpha}{M^2} \kappa^2
\end{equation}
where $\nu_0$ is the pion photoproduction threshold and $\alpha$ the fine-structure
constant.
In a similar way, the forward spin polarizability $\gamma_0$ can
be expressed as:
\begin{equation}\label{eq2}
\gamma_0 = -\frac{1}{4\pi^2}
\int_{\nu_0}^\infty (\sigma_{3/2} - \sigma_{1/2})\,{\hbox{d}\nu \over \nu^3}\ .
\end{equation}

The GDH sum rule, formulated in the 1960's, rests upon basic
physics principles (Lorentz-invariance, gauge invariance,
unitarity) and an unsubtracted dispersion relation applied to the
forward Compton amplitude. Due to its fundamental character this
prediction requires an experimental verification which has been
awaiting technical developments that have only recently been
achieved.

In the absence of any direct experimental result
some theoretical predictions of the GDH integral have been made in the
past years.
They are compared to the GDH sum rule in Table~\ref{tab_pred}.
All predictions were based on  multipole analyses of the existing
single pion photoproduction data
(mainly unpolarized differential and total cross sections and some single polarization observables)
and included an evaluation of the
contribution of double pion photoproduction processes.
With  the exception of Ref.~\cite{burk}, this contribution
has been taken from Ref.~\cite{karli}.
In Ref.~\cite{bb},
an additional contribution due to multi-hadron production processes
was phenomenologically evaluated using a  Regge type approach.

Apart from Ref.~\cite{bb}, all of the theoretical predictions
consistently exceed the sum rule value for the proton while
for the neutron they come out much smaller than the sum rule.
However,  the GDH integrand is an oscillating
function of photon energy, due to multipole contributions of alternating
sign. Therefore, as pointed out in Ref.~\cite{drec},
 a reliable prediction requires a very high accuracy
that has probably not
been reached by any of the existing models, in particular for the neutron.
Precise experimental data are required to pin down the detailed behaviour of the
GDH integrand.
\section{Experimental setup}

The experimental setup has been described previously
in detail \cite{first,MAC96}
and only the main characteristics are given here.
The experiment was carried out at the tagged photon facility~\cite{tagg} of the
MAMI accelerator in Mainz.
Circularly polarized photons were produced by brems\-strahlung of
longitudinally polarized electrons~\cite{aul}.
The electron polarization  (routinely about $75\%$)
was monitored during the data taking by means of
a M{\o}ller polarimeter with a precision of $3\%$.

The photon energy was determined by the Glasgow-Mainz tagging spectrometer
having an energy  resolution of about 2~MeV~\cite{tagg}.
The tagging efficiency was monitored
with an accuracy of $2\%$ by an $e^+e^-$ detector placed
downstream of the main hadron detector\cite{MAC96}.

 Longitudinally polarized protons were provided by a frozen-spin butanol (C$_4$H$_9$OH)
 target~\cite{targ}.
The proton polarization was measured using NMR techniques with a precision
of $1.6\%$,  the target density was known with a precision of $1\%$. 
 Maximum polarization  values close to 90\% were obtained
with a typical relaxation  time of $\simeq$ 200 hours.

Photoemitted  hadrons were registered by a $\approx 4\pi$~sr system based on the
large acceptance detector DAPHNE~\cite{dap}
complemented by detectors~\cite{midas,star} to extend the forward polar angle acceptance.

DAPHNE is essentially a charged particle tracking detector with
cylindrical symmetry.  An outer double scintillator-absorber
sandwich also allows the detection of neutral pions with a
moderate ($15-20\%$) efficiency. It covers polar angles from
$\vartheta_{lab}=21^\circ$ to  $\vartheta_{lab}=159^\circ$.

\section{Data analysis}
In this paper, data recorded by the DAPHNE detector  only are
presented.
An inclusive method of data analysis has been developed  to determine the
total absorption cross section.
It has already been applied to an unpolarized measurement.
Since its features are described in \cite{MAC96,bob}, only the general
characteristics will be recalled here.

A large fraction of the total photoabsorption
cross section ($\sigma_{tot}$)
can be directly accessed by measuring the number of events
with charged hadrons in the final state ($N_{ch}$) detected inside the
DAPHNE acceptance.
Most of the remaining part is deduced from the measured number
of $\pi^0$ events, with no accompanying charged particle observed
($N_{\pi^0}$), by using
the $\pi^0$ detection efficiency ($\overline{\epsilon}_{\pi^0}$)
evaluated by a simulation.
Since  $\overline{\epsilon}_{\pi^0}$ is finite for all $\pi^0$
energies and angles, no extrapolation is needed for the partial
channels having at least one neutral pion in the final state. 
A small correction ($\Delta N_{\pi^0\pi^0,\eta}$) has to be made
since processes involving more than one  $\pi^0$ in the final state
are not included in the evaluation of $\overline{\epsilon}_{\pi^0}$.
An additional correction  $(\Delta N_{\pi^{\pm}})$  is needed
to take into account  the fraction of the events  from the
charged pion channels ($n\pi^+$, $p\pi^+\pi^-$)
emitted  into the angular and momentum regions outside of the detector acceptance.
For E$_\gamma >$ 200 MeV, the region for which the data are presented,
the lower momentum limit for the photoemitted $\pi^+$ from the
$n\pi^+$ reaction is above the detection threshold. Therefore,
no correction for the losses due to the detector momentum acceptance  
is necessary for this channel.

Using the  notation above, $\sigma_{tot}$  can be written as:
\[
\sigma_{tot} \propto N_{ch} + N_{\pi^0}\cdot
 (\overline{\epsilon}_{\pi^0})^{-1}+ \Delta N_{\pi^0\pi^0,\eta} 
+\Delta N_{\pi^{\pm}}.
\]

\begin{figure}[htb]
\begin{center}
\mbox{
\epsfig{file=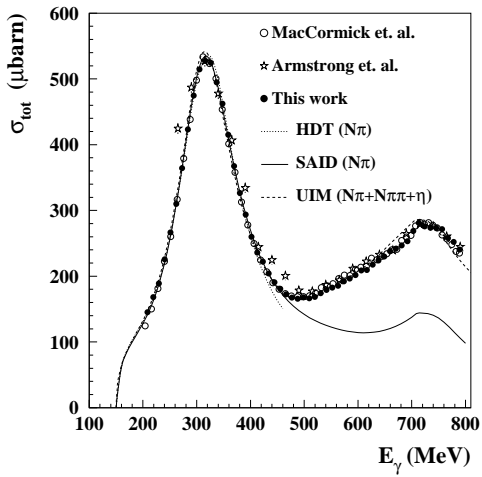,height=7.0cm}
}
\caption{The unpolarized total photoabsorption cross section
on $^1$H obtained in this work is compared to previous
results~\protect\cite{MAC96} (open circles), \protect\cite{arm} (stars)
and to the HDT~\protect\cite{hdt}, SAID~\protect\cite{SAID}
and UIM~\protect\cite{UIM} analyses.
The statistical error bars are smaller than the size of the symbols.
}
\label{stot_unpol}
\end{center}
\end{figure}

In Fig.~\ref{stot_unpol}, the values of $\sigma_{tot}$
obtained with an unpolarized liquid hydrogen target
during a test run, carried out prior to the main experiment,
are shown.
In this case,  $\Delta N_{\pi^{\pm}}$
 (about 5\% of $\sigma_{tot}$) was
evaluated starting from the experimental single charged pion spectra.
The total systematic error on the data is $\simeq \pm 4\%$ of $\sigma_{tot}$.

In the same figure, our data are
compared to previous results~\cite{MAC96,arm} and
to the  HDT prediction from Hanstein {et al.}~\cite{hdt}
 (up to E$_\gamma$ = 450 MeV), the SAID~\cite{SAID} multipole analysis
 and the Unitary Isobar model
(UIM)~\cite{UIM}.
The UIM model takes into account double pion and $\eta$ photoproduction
in addition to single pion production.
The good agreement found with the old data and with the different
models, demonstrates that the detector
response is well understood.

In the analysis of  the data from the polarized butanol target,
the background contribution from the reactions produced on the
unpolarized C and O nuclei of the target could not be fully separated
from the polarized H contribution~\cite{first}.
The inclusive method was then used
to evaluate the difference $\Delta\sigma=$ $(\sigma_{3/2}
- \sigma_{1/2})$ of the two helicity dependent total cross sections
$\sigma_{3/2}$ and $\sigma_{1/2}$.
However,  the extrapolation
of the charged pion photoproduction channels could not be done
as in the unpolarized case. In the following, we briefly describe
the procedures that were used.

The angular extrapolation needed for the $(n\pi^+)$ channel was
evaluated using the SAID~\cite{SAID} multipole analysis, which
reproduced well our previous experimental results of the helicity
dependence of this reaction channel in the $\Delta (1232)$ resonance
region~\cite{first}. 
As a systematic error, $\pm 5\%$ of the evaluated correction
is taken for E$_\gamma < 500$ MeV, while at higher energies,  
$\pm 20\%$ of the evaluated correction is
assumed. 

The correction for the unmeasured part of the  $p\pi^+\pi^-$ channel was
evaluated under the assumption that the  helicity asymmetry
$(\sigma_{3/2}-\sigma_{1/2})/(\sigma_{3/2}+  \sigma_{1/2})$
in the unmeasured part is the same as the one measured inside the
DAPHNE acceptance~\cite{lang}.
The correction for the $p\pi^0\pi^0$ channel
was evaluated similarly~\cite{frank}.
For the $p\eta$ channel, which couples to the $S_{11}$ resonance and
therefore has a dominant $\sigma_{1/2}$ contribution, an helicity
asymmetry of   $-0.97$ was assumed, following the calculation of Ref.~\cite{tiator}.
For the last three contributions,
a systematic error equal to $\pm 50\%$ of the evaluated
correction is assumed.

The remaining sources of systematic errors are due to uncertainties in
wire chamber efficiency (1\% of  $N_{ch}$) and $\pi^0$ detection
efficiency ($4\%$ of $N_{\pi^0}\cdot (\overline{\epsilon}_{\pi^0})^{-1})$.
The dominant contribution to the systematic error stems from uncertainties in
photon flux, target density and beam and target polarizations; their
sum in quadrature is about $4\%$ of $\Delta \sigma$.
Tab.~\ref{tabsyserr} summarizes for two energies
the different contributions to $\Delta\sigma$ together with their
systematic errors and gives the total  systematic error
obtained by summing in quadrature all contributions.

\section{Results}
\begin{figure}[htb]
\begin{center}
\mbox{
\epsfig{file=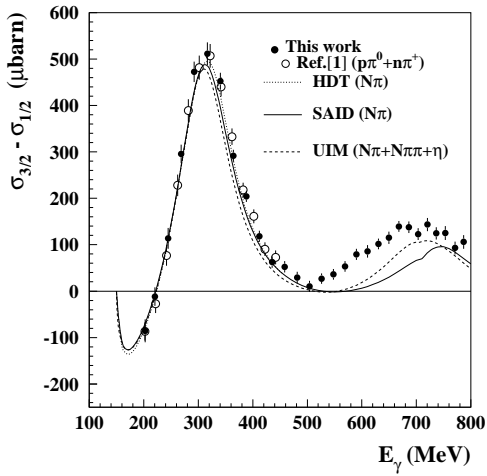,height=7.0cm}
}
\caption{The total cross section difference $(\sigma_{3/2}-\sigma_{1/2})$
on $^1$H is compared to previous results~\protect\cite{first}
(open circles) and to the predictions of the HDT~\protect\cite{hdt}, 
SAID~\protect\cite{SAID} and UIM~\protect\cite{UIM} analyses.
Only statistical errors are shown.
}
\label{stot_pol}
\end{center}
\end{figure}

The analysis procedure as described above results in the total cross section
difference $(\sigma_{3/2}-\sigma_{1/2})$  depicted
in Fig.~\ref{stot_pol}~\cite{bob}.
It is compared with the sum of our previously published
helicity differences for the $n\pi^+$ and $p\pi^0$
channels in the $\Delta$ region~\cite{first}.
The good agreement found between the different analyses gives us confidence
in their reliabilty.
In the same figure, our data are also compared
to the  HDT~\cite{hdt}, SAID~\cite{SAID},
and UIM~\cite{UIM} analyses.

In the $\Delta$ resonance region, there is a rather good agreement
between experiment and theories.
In the second resonance region, a significant contribution
from double pion photoproduction is clearly visible.
This feature is not completely reproduced by the UIM  model.

\begin{figure}[htb]
\begin{center}
\mbox{
\epsfig{file=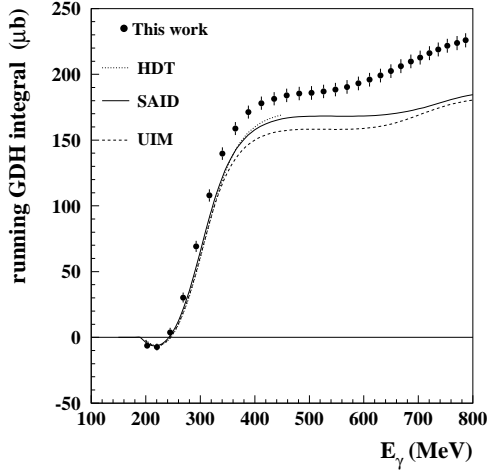,height=7.0cm}
}
\caption{The running GDH integral obtained in this work starting at 200 MeV is compared to the
model predictions. Only statistical errors are shown.
}
\label{integ_gdh}
\end{center}
\end{figure}

In Fig.~\ref{integ_gdh} the experimental running  GDH integral
(left-hand side of eq.\ref{eq1}) is displayed and compared
to the model predictions. The integration starts at E$_\gamma = 200$~MeV
and the upper integration limit is taken as the running variable.
The measured value of the GDH integral between 200 and 800 MeV amounts to
$226\pm 5 ~(stat) \pm 12~(sys) ~\mu$b.

Due to the $\nu^{-3}$ weighting, the
 $\gamma_0$  running integral  is almost saturated
by E$_\gamma = 800$~MeV.
The  value of the $\gamma_0$ integral
between 200 and 800 MeV amounts to
$[-187\pm 8 ~(stat) \pm 10 ~(sys)]\cdot 10^{-6}~\hbox{\rm fm}^4$.

\section{Discussion}

Although the measured photon energy interval is too narrow to
draw any definitive conclusion,  a reasonable estimate of the GDH
sum rule  value can be deduced if we use the existing models for
the evaluation of the missing contributions. The UIM model
\cite{UIM} gives a contribution of $-30 \mu$b for  E$_\gamma <
200$~MeV and $+40 \mu$b for $800 <$ E$_\gamma < 1650$~MeV. For
E$_\gamma > 1650$~MeV, Ref. \cite{bb} gives a contribution of -26
$\mu$b. The combination of our experimental result with these
predictions yields an estimate (210 $\mu$b) which within the
experimental errors is consistent with the GDH sum rule value
(\ref{eq1}). It should be kept in mind that, especially above
E$_\gamma = 800$~MeV, none of the models has yet been validated
experimentally and only a measurement in this energy region can
lead to a definitive conclusion about the high energy
contribution to the GDH-integral. Our collaboration  is
performing such a measurement at ELSA (Bonn) up to E$_\gamma
\simeq 3$~GeV, and extensions to higher energies are under way or
in preparation at Jefferson Lab~\cite{sober}  and
SLAC~\cite{bosted}.

In case of the $\gamma_0$-integral (\ref{eq2}) the contribution from
E$_\gamma < 200$~MeV is important, the UIM prediction being
$+104 \cdot 10^{-6}$ fm$^4$.
The missing high energy contribution, according to UIM and Ref. \cite{bb},
is $-3\cdot10^{-6}$ fm$^4$ only. The combination with our experimental result
gives an estimate of $-86\cdot 10^{-6}$ fm$^4$ for $\gamma_0$ .

Several predictions, based on dispersion relations \cite{BGLM,drec2}
and chiral perturbation theory \cite{BKM95,hem98,JKO,KMB,GHM},
have been made for $\gamma_0$  in the last few years, see table \ref{tab_spin}.
They range from $(-380$ to $+460$) $\cdot10^{-6}\hbox{ fm}^4$. Our result
is close to the range of $\gamma_0$  values predicted by dispersion theory.

The authors wish to acknowledge the excellent support of the accelerator group of MAMI.
We also thank D.~Drechsel and his group for
their important contribution to the interpretation of the data.
This work was supported by
the Deutsche Forschungsgemeinschaft (SFB 201, SFB 443 and Schwerpunktprogramm), the
INFN--Italy,  the FWO Vlaanderen--Belgium, the IWT--Belgium,
the UK Engineering and Physical Science Council, the DAAD, and the Grant-in-Aid,
Monbusho, Japan.

\def\Journal#1#2#3#4{{#1} {\bf #2}, {#3} (#4)}

\def\NCA{Nuovo Cimento}
\def\NIMA{{Nucl. Instrum. Methods Phys. Res. Sect. A}}
\def\NPA{{Nucl. Phys.} {\bf{A}}}
\def\NPB{{Nucl. Phys.} {\bf{B}}}
\def\PLB{{Phys. Lett.} {\bf{B}}}
\def\PRL{Phys. Rev. Lett.}
\def\PRD{{Phys. Rev.} D}
\def\PRC{{Phys. Rev.} C}
\def\PPNP{{Prog. Part. Nucl. Phys.}}
\def\SJNP{{ Sov. J. Nucl. Phys.}}
\def\IJMP{Int. J. Mod. Phys.}
%

%
%

%
%
%
%
%
%

%
%
%
\begin{table}[htbb]
\caption{The GDH sum rule (eq. \ref{eq1}) for the proton ($I^{GDH}_p$) and the neutron ($I^{GDH}_n$)
is compared with some theoretical predictions.}
\label{tab_pred}
\begin{tabular}{|l|c|c|c|}
     &  {$I_{p}^{GDH}$} &   {$I_{n}^{GDH}$} &   {$I_{(p-n)}^{GDH}$}    \\
 &(${\mu b}$) &(${\mu b}$) & (${\mu b}$) \\[0.1cm] \hline
GDH sum rule & { 205} &  { 233} &  { - 28 }\\[0.2cm]\hline\hline
Karliner~\cite{karli} &   261   &  183    &    78    \\
Workman-Arndt~\cite{a}&   260   &  157    &    68    \\
Burkert-Li~\cite{burk}&   223   &         &         \\
Sandorfi et al.~\cite{s} & 289   &  160    &   129    \\
Drechsel-Krein~\cite{drec}& 261&  180    &    81    \\
Bianchi-Thomas~\cite{bb}  & 207$\pm$ 23   &  226 $\pm$ 22  &   -19 $\pm$37 \\
\end{tabular}
\end{table}
%
%
%
\begin{table}
\caption{
The different contributions (in $\mu$b) to $\Delta\sigma$ and to the total
systematic error are shown at E$_{\gamma}$=317~MeV and
E$_{\gamma}$=753~MeV. The symbol $\delta(t,\gamma)$ denotes
the sum in quadrature of the systematic errors related
to photon flux, beam and target polarizations, and target
density.  See the text for the explanation of the other symbols.
}
\label{tabsyserr}
\begin{tabular}{|c|c|c|c|c|c|c|c|}
E$_\gamma$ & $\Delta\sigma $ &  $ N_{ch}$ &
$N_{\pi^0}\cdot(\overline{\epsilon}_{\pi^0})^{-1}$ & $\Delta N_{\pi^{\pm}}$
& $\Delta N_{\pi^0\pi^0,\eta}$ & $\delta(t,\gamma)$ & total  \\
(MeV)      &  ($\mu$b)      &  &  &   &  & & sys. err. \\
\hline
317 & 511 & 315 $\pm$ 3 & 213 $\pm$ 9 &  -17 $\pm$ 1 & & 21 &23 $\mu$b \\
753 & 125 & 107 $\pm$ 1 & 22  $\pm$ 1 &  -7 $\pm$ 4 & 3 $\pm$ 2 & 5  & 7 $\mu$b
\\
\end{tabular}
\end{table}
%
\begin{table}
\caption{Theoretical predictions of the forward spin
polarizability $\gamma_0$ of the proton (in units of 10$^{-6}$fm$^{4}$).}

\label{tab_spin}
\begin{tabular}{|l|l|l|l|}
Dispersion theory& $\gamma_0$ &  ChPT&  $\gamma_0$ \\[0.2cm]\hline\hline
Babusci et al.~\cite{BGLM}&  -150     &  Bernard et al.~\cite{BKM95}&   +460    \\
Drechsel et al.~\cite{drec2}&      -80         &  Hemmert et al.~\cite{hem98}&     +200   \\
                     &      &  Ji et al.~\cite{JKO}&   -380     \\
                     &      &  Kumar et al.~\cite{KMB}&  -380    \\
                     &      &  Gellas et al.~\cite{GHM}& -100       \\
\end{tabular}
\end{table}

\end{document}